\documentclass[twocolumn]{article}

\usepackage{dcolumn}
\usepackage{bm}
\usepackage{lipsum}

\usepackage[pdftex]{graphicx}
\usepackage[margin=25truemm]{geometry}

\title{Novel Slow Dynamics of Phase Transition\\ in the Partially Ordered Frustrated Magnet $\mathrm{DyRu_2Si_2}$}

\author{Subaru Yoshimoto, Yoshikazu Tabata, Takeshi Waki, and Hiroyuki Nakamura\\ 
\textit{Department of Materials Science and Engineering, Kyoto University, Kyoto 606-8501, Japan}} 
\date{}

\begin{document}
\maketitle

\begin{abstract}
$\mathrm{DyRu_2Si_2}$ is a frustrated magnet exhibiting multiple magnetic phase transition in zero and finite magnetic fields.
 We investigated and characterized the phase transition between
  the partially-ordered antiferromagnetic phases at zero field by ac susceptibility measurements.
   Detailed ac susceptibility measurements reveal the novel critical dynamics of the phase transition;
    extremely slow dynamics with the relaxation time 
    in the order of 10-100 ms,
    speed-up of the dynamics on
    cooling indicating its non-thermally activated origin
    and growing of the ferromagnetic correlations towards the phase transition temperature.
    On the basis of these findings, we propose a novel  phase transition process, namely, the spontaneous 
    striped-arrangement of the precedently emergent ``belt-like'' ferromagnetic spin textures.

\end{abstract}

\section{Introduction}

 Frustrated magnets have been widely investigated for their rich variety of properties.
 In a frustrated system, where interactions compete,
 a vast number of physical states having the same energy coexist and, therefore, the system is highly degenerated. 
 These frustrated systems, such as spin glass and spin ice, are known to exhibit slow and complicated dynamics at low temperature 
 because of their high degeneracy\cite{spinglassslowdynamics, spiniceslowdynamics, spinglassKG, spinglassPG, spinglassKJ, spiniceJAQ, spiniceHT}.
   \par
   
   In the present report, we focus on the novel slow dynamics in the frustrated magnet 
  $\mathrm{DyRu_2Si_2}$.
  It is one of the series of the intermetallic compound
 $\mathit{RT_\mathrm{2} X_\mathrm{2}}$
 ($\mathit{R} =$ rare earth, 
  $\mathit{T} =$ 4d or 5d metal, 
  $\mathit{X}$ = $\mathrm{Si}$ or $\mathrm{Ge}$), with the tetragonal
$\mathrm{ThCr_2Si_2}$-type structure, which exhibits a diversity of magnetic properties such as ferromagnetism, 
  antiferromagnetism and paramagnetic heavy fermion
\cite{R=NdTb,R=Ce,R=Pr}. 
   The magnetic properties of 
    $\mathit{RT_\mathrm{2} X_\mathrm{2}}$ are controlled by the interaction between the conduction electrons and f-electrons.
   In the heavy rare earth compounds, the f-electrons have strongly localized nature and complicated frustrated magnetism is
   observed\cite{DyRu2Ge2Multistep,R=NdTb} due to the frustration effect of the oscillating long-range RKKY interaction.\par

\begin{figure}
\vspace*{1.5cm}
\centering
\includegraphics[width=\columnwidth]
{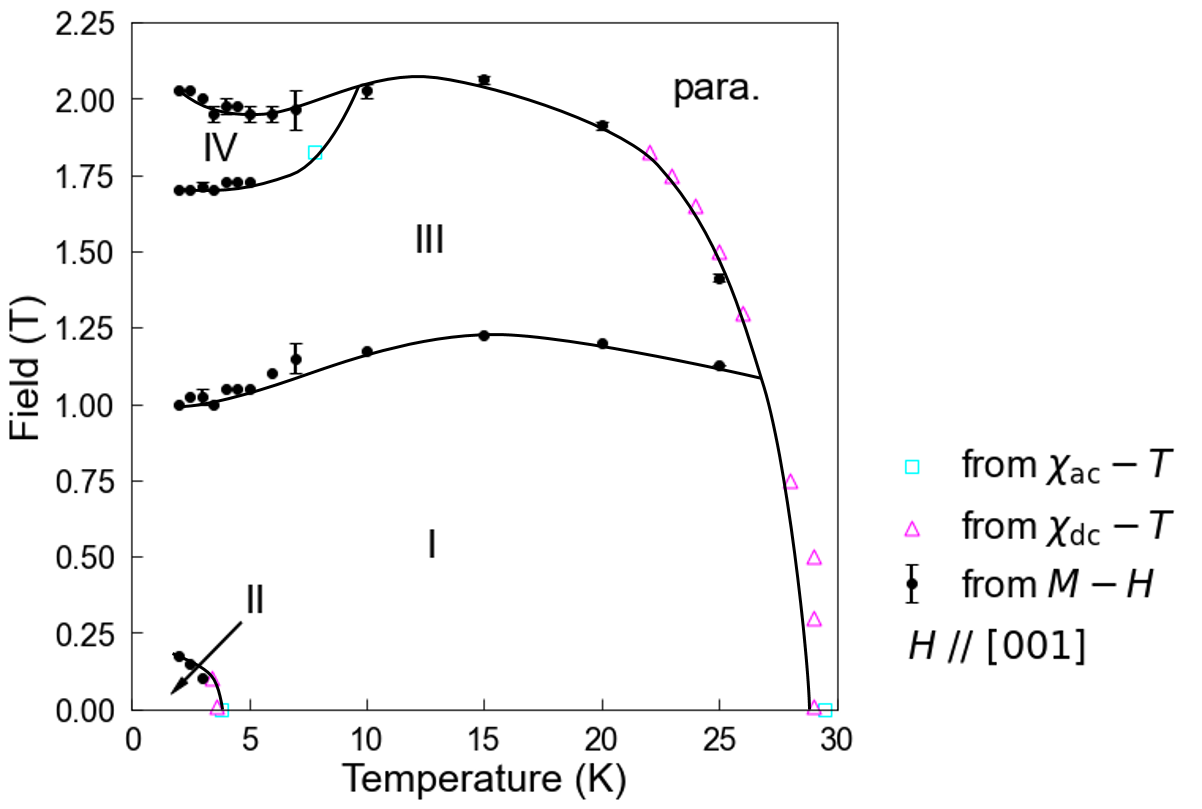}
\caption{(Color online) $H \mathchar`- T$ phase diagram of $\mathrm{DyRu_2Si_2}$. This phase diagram was reported in the literature\cite{DyRu2Si2Properties} and confirmed in the present study.}
\label{phase_diag}
\end{figure}

\begin{figure*}[t]
\vspace*{1.5cm}
	\begin{tabular}{cc}
	\centering
		\begin{minipage}{0.5\linewidth}
    		\centering
    		\includegraphics[scale=0.5]{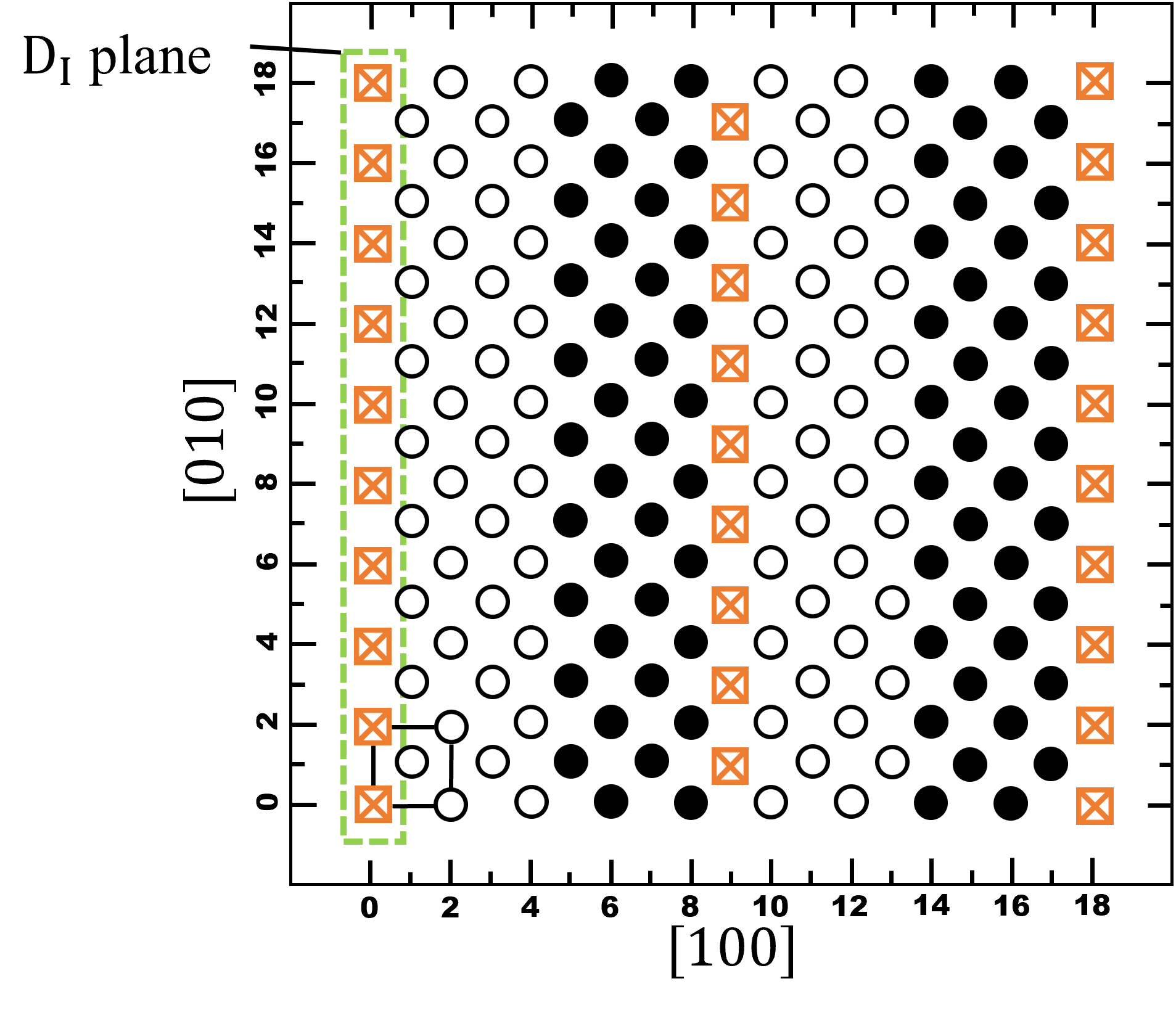}
  		\end{minipage}&
  		
  		\hspace{-1cm}
  
		\begin{minipage}{0.5\linewidth}
    		\centering
    		\includegraphics[scale=0.5]{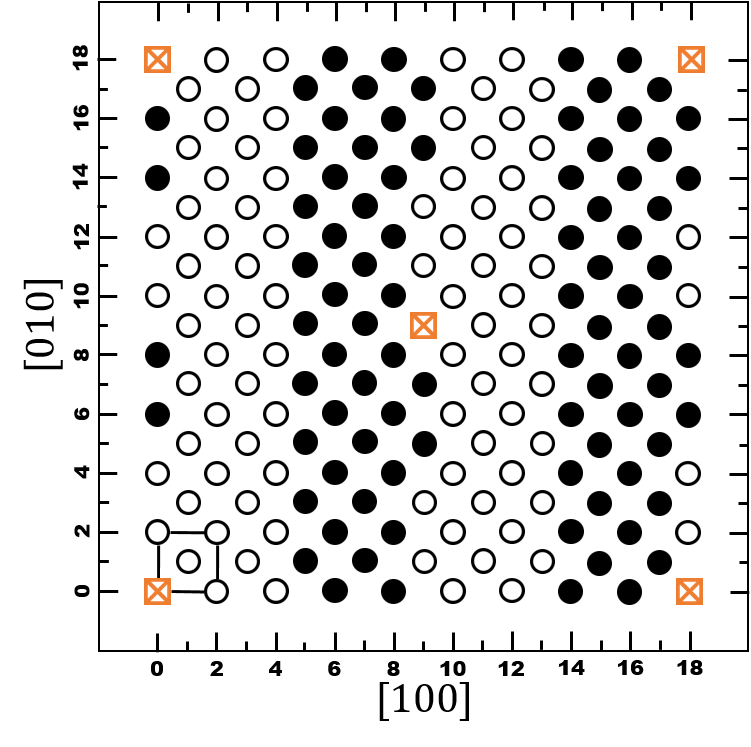}
		\end{minipage}\\
		
		\begin{minipage}{0.5\linewidth}
		\centering
		(a)
		\end{minipage}&
		
		\hspace{-1cm}
		
		\begin{minipage}{0.5\linewidth}
		\centering
		(b)
		\end{minipage}
		
  	\end{tabular}
  \caption{(Color online) Schematic views of the magnetic structures projected on the basal c-plane of (a) the phase I and (b) II of 
  $\mathrm{DyRu_2Si_2}$.
  A square on the left bottom of each figure represents the crystallographic unit cell and each circle or square represents a $\mathrm{Dy}$ ion on the corner or body-centered position. 
  The unit of axes is half a lattice constant $a/2$. 
  The open and closed circles indicate the Dy ions with the spins parallel and antiparallel to the c-axis, respectively. 
  The orange square with a cross represents the Dy ion with the fluctuating spin. }
  \label{magn_struc}
\end{figure*}

   Among them, 
   $\mathrm{DyRu_2Si_2}$ is a representative frustrated 
    $\mathit{RT_\mathrm{2} X_\mathrm{2}}$, where magnetic $\mathrm{Dy^{3+}}$ ions
   have a strong c-axis anisotropy\cite{DyRu2Si2Properties}.
  It exhibits multistep phase transition against temperature and magnetic field and has a complicated 
  $H \mathchar`- T$ phase diagram as shown in Fig. \ref{phase_diag}, which has paramagnetic and four antiferromagnetic phases. 
  Kawano et al. have revealed the magnetic structure of each phase from the results of neutron scattering\cite{DyRu2Si2MagneticStructure}. 
  In zero magnetic field, it was indicated that the phase I and II are partially antiferromagnetically ordered phases, where fluctuating spins still remain even below their transition temperatures.
 The magnetic structures of these phases, which are extensively investigated in this study, are schematically shown in 
 Figs.\  \ref{magn_struc} (a) and (b). 
 The phase I has a stripe structure which has a long period along the a-axis with the propagation vector $Q = (2/9, 0, 0)$\cite{DyRu2Si2MagneticStructure}.
   Considering the $C_4$ symmetry of the crystal structure, one expects the formation of two equivalent domains A and B, which have the period along the a- and b-axes, respectively.
   However, the magnetic responses of two domains against the magnetic field along the c-axis are indistinguishable, and thus all description hereafter is based on only the A-domain for simplicity.
 It is noteworthy that disordered paramagnetic a-planes (denoted as the
   $\mathrm{D_I}$ planes hereafter) appear every 9 ordered a-planes.
   It is indicated that the spins in the 
   $\mathrm{D_I}$ plane are magnetically free, namely, interactions from the neighboring and further ordered a-planes compete with each other
   and are canceled out.
   Thus, these spins on the 
   $\mathrm{D_I}$ planes can be considered as a pseudo two-dimensional system. 
   The phase II has the magnetic structure where the  spins in the
   $\mathrm{D_I}$ planes are partly ordered along the b-axis. 
Such partially ordered states due to the insufficient lift of the degeneracy are also found in other frustrated magnets
 \cite{Na3CoCoexistingMagneticOrder,CePdAl,Gd2Ti2O72kOrder,
 Gd2Ti2O7,Gd2Ti2O7PartialOrder,Sr2CoOsO6SpinDynamics,GdTi2O7Multi-kStructure,UNi4B}.
    In these systems, slow spin dynamics owing to the paramagnetic, but strongly correlated, fragment is often
     observed\cite{YCuKagomePartialOrder}.    
   Thus, from the magnetic structures of the partially ordered phases I and II, 
   one can expect novel slow dynamics in this 
   $\mathrm{DyRu_2Si_2}$ due to the fluctuating spins. 

   In the present study, we have performed detailed ac-susceptibility measurements to reveal it and have found novel critical dynamics accompanying the phase transition between these partially ordered phases I and II. 
   The striking features of the novel dynamics are the following three points. 
   First, its relaxation time is extremely long (order of 10-100 ms).
   Second, the dynamics becomes faster on cooling indicating non-thermally activated origin. 
   Third, ferromagnetic correlations grow towards the phase transition temperature even though both phases are antiferromagnetic.
   Of course, in spin glass or spin ice, slow dynamics with the relaxation time of s-ms order is often observed\cite{spinglassslowdynamics, spiniceslowdynamics}.
    In such a system, it can be attributed to the glassy nature due to the high degeneracy at low temperature.
   However, since this is not the case in 
   $\mathrm{DyRu_2Si_2}$, it is an intriguing outcome that we observed such a long relaxation time over the I-II phase transition.
   On the basis of these findings, we proposed the process of the phase transition, namely, the spontaneous stripe-arrangement of the precedently
   emergent ``belt-like'' ferromagnetic spin textures.

\begin{figure}[h]
\centering
\includegraphics[width=7cm,bb=0 0 576 720]{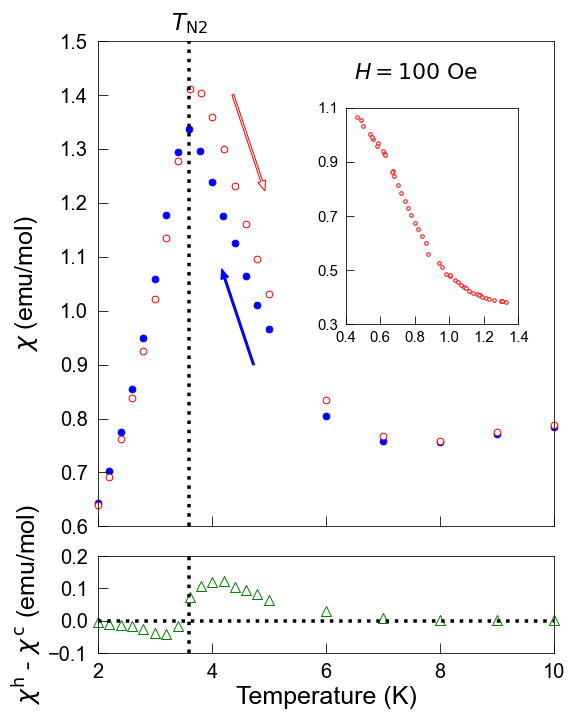}
\caption{(Color online) Temperature hysteresis of the dc susceptibility over the I-II phase transition.
			$\chi^\mathrm{h}$ and $\chi^\mathrm{c}$ are denoted as red-open and blue-closed circles, respectively. 
			Green-open triangles represent the difference of $\chi^\mathrm{h}$ and $\chi^\mathrm{c}$.
			The inset shows the susceptibility in the temperature region below the I-II phase transition.
			The measurement protocols are described in the text.
			}
\label{T_hys}
\end{figure}

\section{Experimental Details}

\begin{figure*}[!t]
\vspace*{0cm}
	\begin{tabular}{cc}
	\centering
		\begin{minipage}{0.5\linewidth}
    		\centering
    		\includegraphics[scale=0.35, bb = 0 0 720 936]{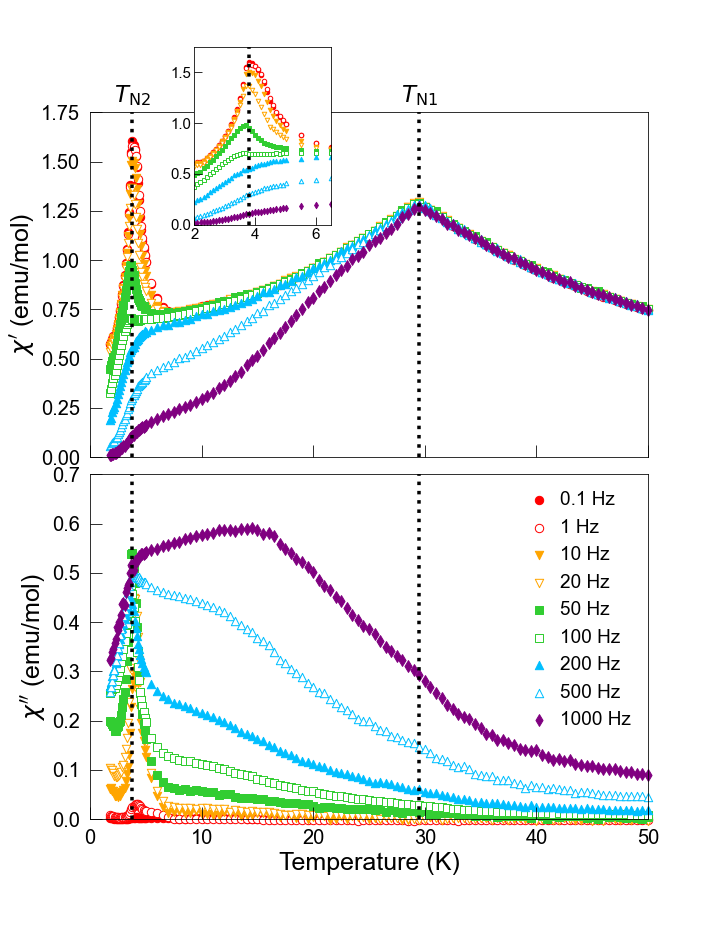}
  		\end{minipage}&
  
		\begin{minipage}{0.5\linewidth}
    		\centering
    		\includegraphics[scale=0.35, bb = 0 0 720 936]{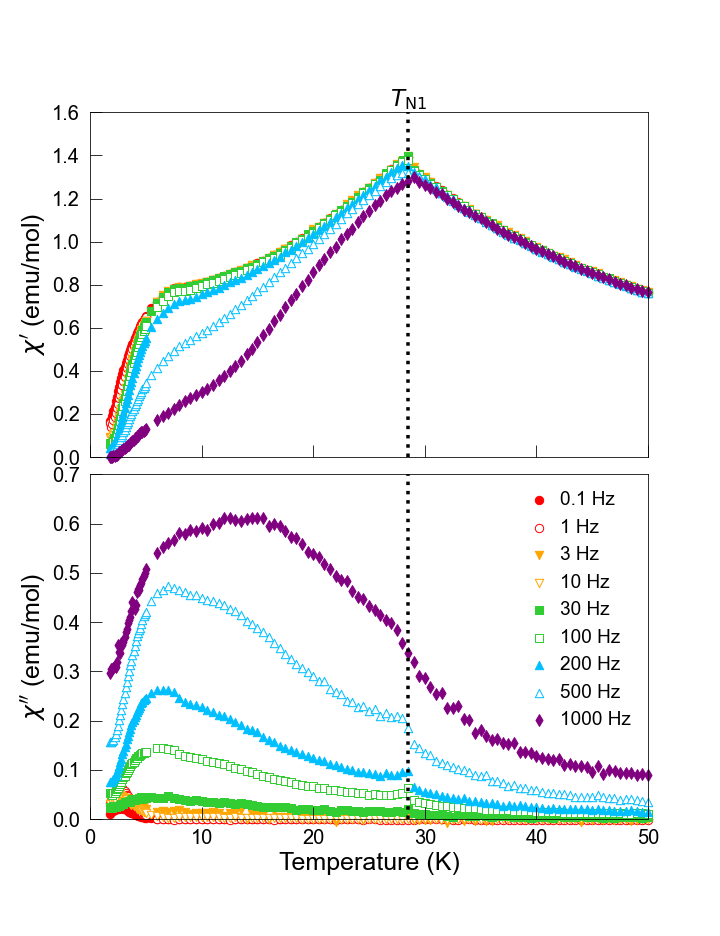}
		\end{minipage}\\
		
		\begin{minipage}{0.5\linewidth}
		\centering
		(a)
		\end{minipage}&
		
		\begin{minipage}{0.5\linewidth}
		\centering
		(b)
		\end{minipage}
		
  	\end{tabular}
  \caption{(Color online) Temperature dependences of the real (upper) and imaginary (bottom) parts of the ac susceptibility at the bias field of (a) 0 and 
  (b) $5\ \mathrm{kOe}$. The inset of the upper figure in (a) is the enlarged figure of the low-temperature region.
    Both temperature dependences were measured with elevating temperature after ZFC.}
  \label{chi_ac-T}
\end{figure*}

We synthesized a polycrystalline sample with an arc furnace followed 
by single crystal growth by the Czochralski pulling method with a tetra arc furnace. 
The grown single crystal was cut, and finally, a cubic-like-shaped sample with the weight of 
$10.1\ \mathrm{mg}$ was obtained for dc and ac susceptibility measurements.
We performed dc and ac susceptibility measurements using the SQUID
  magnetometer (MPMS, Quantum Design) equipped in the Research Center 
  for Low Temperature and Materials Sciences, Kyoto University.
  Firstly, the magnetic field dependences of magnetization at several temperatures 
  and the temperature dependences of the dc and ac susceptibilities at several magnetic fields were examined to confirm the 
  $H \mathchar`- T$ phase diagram of 
  $\mathrm{DyRu_2Si_2}$ in the literature\cite{DyRu2Si2Properties}. 
  The result is denoted in Fig. \ref{phase_diag}.
  Also, we measured detailed frequency dependences of ac susceptibility 
  in the vicinity of the I-II phase transition temperature 
  at zero bias field for revealing the critical dynamics and at 
  $H = 5\  \mathrm{kOe}$ for a comparison.
  The ac susceptibility  measurements were performed with the amplitude of the oscillating field of
  $3 \ \mathrm{Oe}$ and the frequency region of 
  $0.1\mathchar`-1000 \ \mathrm{Hz}$.

\section{Results}

  The temperature dependences of the dc susceptibility over the I-II phase transition at 100 Oe in the cooling and heating processes are shown in Fig. \ref{T_hys}.
  On the measurement, we first cooled down to 
  $10\ \mathrm{K}$ under zero-field-cooled (ZFC)
  condition and set magnetic field of 
  $100\ \mathrm{Oe}$. 
  Then we measured magnetization on cooling down to 
  $1.8\ \mathrm{K}$ and went backward.
  Let 
  $\chi^\mathrm{h}$, 
  $\chi^\mathrm{c}$ and 
  $T_\mathrm{N2}$ be the susceptibilities measured on heating and on cooling and 
  the transition temperature of the I-II phase transition, respectively. 
  The temperature dependences of 
  $\chi^\mathrm{h}$ and
  $\chi^\mathrm{c}$ are roughly similar.
   The peak temperature of 
  $\chi^\mathrm{h}$ and
  $\chi^\mathrm{c}$, which are the sign of the phase transition, are the same.
  Although, since there is the hysteresis behavior, the phase transition temperature 
  $T_\mathrm{N2}$ cannot be accurately identified, it is approximately evaluated as 3.6 K.
  This corresponds to the phase transition temperature reported in the earlier work\cite{DyRu2Si2Properties}.
   As shown in the lower panel of Fig. \ref{T_hys}, the magnitude of  
  $\chi^\mathrm{h}$ is greater than that of
  $\chi^\mathrm{c}$ in the temperature region of
  $T \ge T_\mathrm{N2}$ and, vice versa, 
  $\chi^\mathrm{c} > \chi^\mathrm{h}$ for
  $T < T_\mathrm{N2} $.
  The hysteresis behavior of the dc susceptibility indicates the presence of slow dynamics accompanying the I-II phase transition in the time scale of the measurement or longer and was not observed over the para-I phase transition (not shown).

  \begin{figure*}[!t]
\vspace*{0cm}
	\begin{tabular}{cc}
	\centering
		\begin{minipage}[t]{8cm}
    		\centering
    		\includegraphics[scale=0.6, bb = 0 0 400 500]{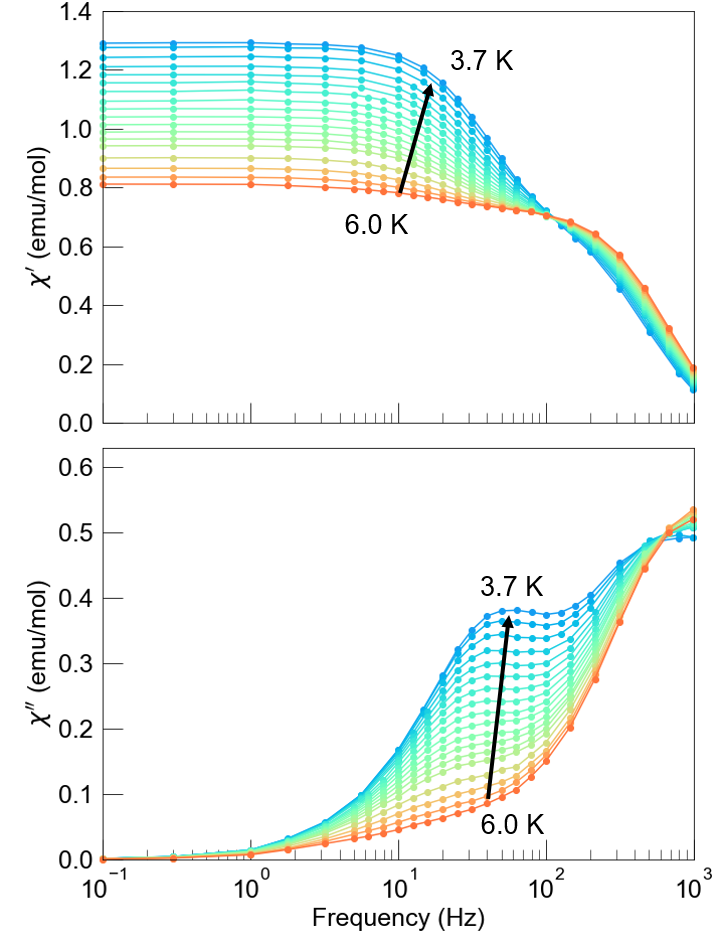}
  		\end{minipage}&
  
		\begin{minipage}[t]{8cm}
    		\centering
    		\includegraphics[scale=0.6, bb = 0 0 400 450]{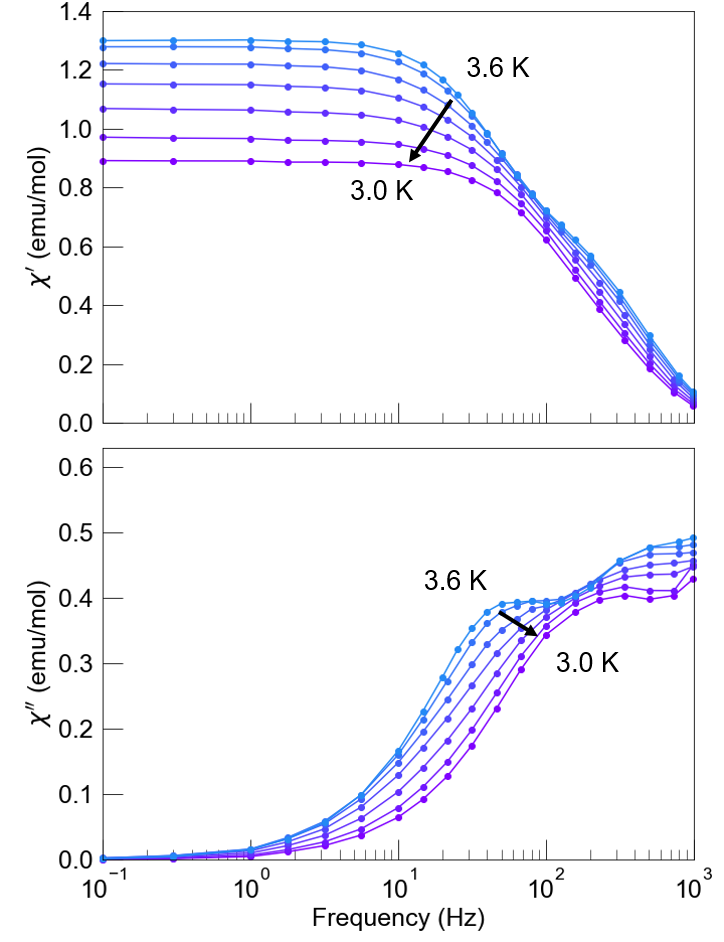}
		\end{minipage}\\
		
		\begin{minipage}{0.5\linewidth}
		\centering
		(a)
		\end{minipage}&
		
		\begin{minipage}{0.5\linewidth}
		\centering
		(b)
		\end{minipage}
		
  	\end{tabular}
  \caption{(Color online) Frequency dependences of the real (upper) and imaginary (lower) parts of the ac susceptibility 
  (a) above $3.7\ \mathrm{K}$ and (b) below $3.6\ \mathrm{K}$ at zero field measured on cooling.}
  \label{chi-f}
\end{figure*}

   The inset of Fig. \ref{T_hys} shows the lower temperature susceptibility below
   $T_\mathrm{N2}$.
   The measurement was performed with the heating process after ZFC down to 0.46 K and applying magnetic field of 100 Oe.
   Here, it is found that the susceptibility increases down to the lowest measured temperature
   $0.46\ \mathrm{K}$, which indicates the presence of the fluctuating Dy spins 
   in the phase II (Fig. \ref{magn_struc} (b)) and its persistence down to this temperature.
   This result is consistent with the increase of the specific heat towards the lowest temperature in the phase II, reported in the previous study\cite{DyRu2Si2Properties}. 
   We are not sure whether there is temperature hysteresis in this temperature region as well as above 1.8 K and whether
   there is another phase transition, where the fluctuating spins order, at lower temperature.
   It will be investigated in the near future.

  Figures \ref{chi_ac-T} (a) and (b) are the temperature dependences of the ac susceptibility, the real part 
  $\chi^\prime$ and imaginary part
  $\chi ^{\prime \prime}$ at the bias field of
   0 and $5\ \mathrm{kOe}$. 
   At zero field (Fig.\ \ref{chi_ac-T} (a)), two-step phase transition at 
   $T_\mathrm{N1}=29.5\ \mathrm{K}$ and at
   $T_\mathrm{N2}$, corresponding to the para-I and I-II phase transitions respectively, are clearly seen.
   Here 
   $T_\mathrm{N1}$ is the phase transition temperature of the para-I phase transition.    
   In the plot of the real part 
   $\chi^\prime$ at zero field, the peak temperature
   corresponding to the I-II phase transition is $3.8\ \mathrm{K}$, which is slightly different from 
   $3.6\ \mathrm{K}$ of the peak temperature in the dc susceptibility.
   This difference should come from the non-equilibrium effect in the vicinity of 
   $T_\mathrm{N2}$.
   The transition temperature should ideally be unique and thus we have to say that the true
   transition temperature is not able to be identified but some value around
   $\mathrm{3.6\ K}$. 
   In zero field, there is no frequency dependence of 
   $\chi^\prime$ in the paramagnetic phase and in the vicinity of the para-I phase transition,
   whereas strong frequency dependence of 
   $\chi^\prime$ and corresponding substantial
   $\chi ^{\prime \prime}$ appear 
   especially above 
   $100\ \mathrm{Hz}$
    in the phase I and II. 
    It is remarkable that the striking frequency dependence of 
    $\chi^\prime$ in the lower frequency region is found in the vicinity of  
    $T_\mathrm{N2}$, where the peak attenuates with increasing frequency and disappears above 100 Hz.
    Correspondingly, a sharp peak of 
    $\chi ^{\prime \prime}$ and its suppression with increasing frequency are observed at 
    $T_\mathrm{N2}$.
     At $H=5\ \mathrm{kOe}$ (Fig. \ref{chi_ac-T} (b)), similar behavior of both 
     $\chi^\prime$ and $\chi ^{\prime \prime}$ are seen in the paramagnetic phase and phase I,
    whereas, the peak anomalies of 
    $\chi^\prime$ and $\chi ^{\prime \prime}$ attributed to the I-II phase transition are absent.
    These results indicate the presence of slow dynamics in the phase I and II with long relaxation times of the order of 10 ms.
    This feature should be attributed to the fluctuating spins in each partially ordered phase.
    The slower dynamics in the vicinity of 
    $T_\mathrm{N2}$ is more striking and should be associated with the characteristic hysteresis behavior of the dc susceptibility.
    \par

  In order to investigate the slow dynamics attributed to the I-II phase transition more deeply,
   we measured more detailed frequency dependences of the ac susceptibility between the temperature of
   $3.0\ \mathrm{K}$ and
   $6.0\ \mathrm{K}$ at zero field.
   Since the I-II phase transition shows the temperature hysteresis, 
   we measured the frequency dependences on both cooling and heating.
   The frequency dependences measured on cooling are shown in Figs. \ref{chi-f} (a) and (b), which show the plots of the temperature range 
   above 3.7 K and below 3.6 K, respectively.
   In Fig. \ref{chi-f} (a), the real part 
   $\chi^\prime$ at 6.0 K shows a one-step-like structure which has a reduction at around 
   $200\ \mathrm{Hz}$ and it changes to a two-step-like structure, where another reduction at around 
   10 Hz appears, with approaching 
   $T_\mathrm{N2}$.
   In Fig. \ref{chi-f} (b), it changes to a one-step-like one again with decreasing temperature further.
    One can see the feature of the change of dynamics more clearly in the imaginary part
    $\chi^{\prime \prime}$.
    At 6.0 K in Fig. \ref{chi-f} (a), it shows a one-peak-like structure with a peak around  
     $1000\ \mathrm{Hz}$, or higher, 
     corresponding to the one-step-like structure of 
     $\chi^\prime$ at 6.0 K.
     With decreasing temperature, an additional peak appears around 10 Hz.
   This peak shifts to a higher frequency with approaching 
   $T_\mathrm{N2}$.
   In Fig. \ref{chi-f} (b), it merges into the higher frequency peak further below 
   $T_\mathrm{N2}$. 
   These changes of 
   $\chi ^{\prime \prime}$ correspond to the development of the two-step-like structure and the retransformation to the one-step-like one in 
   $\chi^\prime$.
   These results indicate that the system has several relaxation components owing to the phases I and II themselves and the I-II phase transition. 
   It is noteworthy that the characteristic frequency, owing to the I-II phase transition that gives the reduction of 
   $\chi^\prime$ and the peak of 
   $\chi ^{\prime \prime}$ observed below 6.0 K, increases with decreasing temperature. 
  The frequency dependences of the ac susceptibility measured on heating exhibits similar behavior, but is greater in magnitude.

     \section{Analysis}

\begin{figure}[t]
\vspace*{1cm}

\centering
\includegraphics[width=8cm,bb=0 0 664 623]{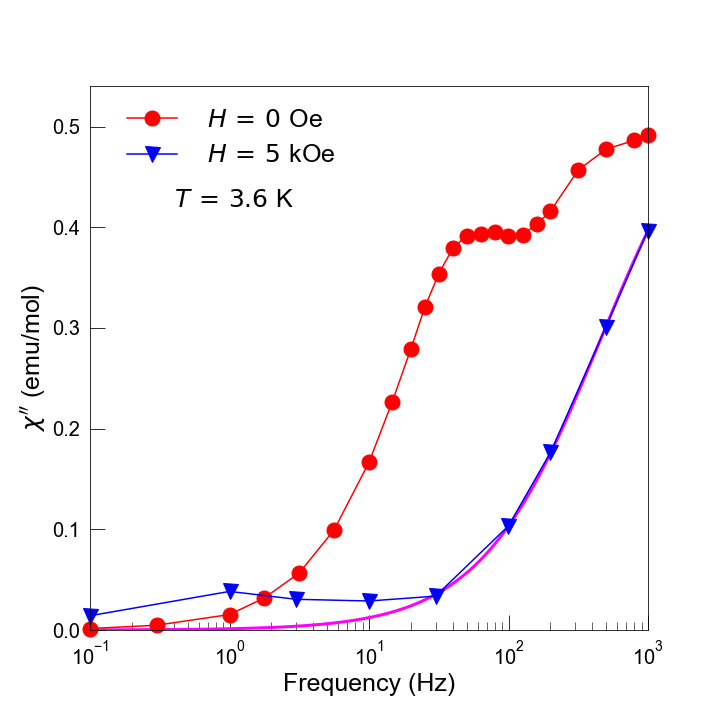}
\caption{(Color online) Comparison of the frequency dependences of 
$\chi ^{\prime \prime}$ in the field of 0 and 5 kOe at 
$3.6 \ \mathrm{K}$ in the cooling process. The pink solid curve is the fitting curve (see the main text). }
\label{compare}
\end{figure}

\begin{figure}[t]
\vspace*{1cm}
\centering
\includegraphics[width=8cm,bb=0 0 664 623]{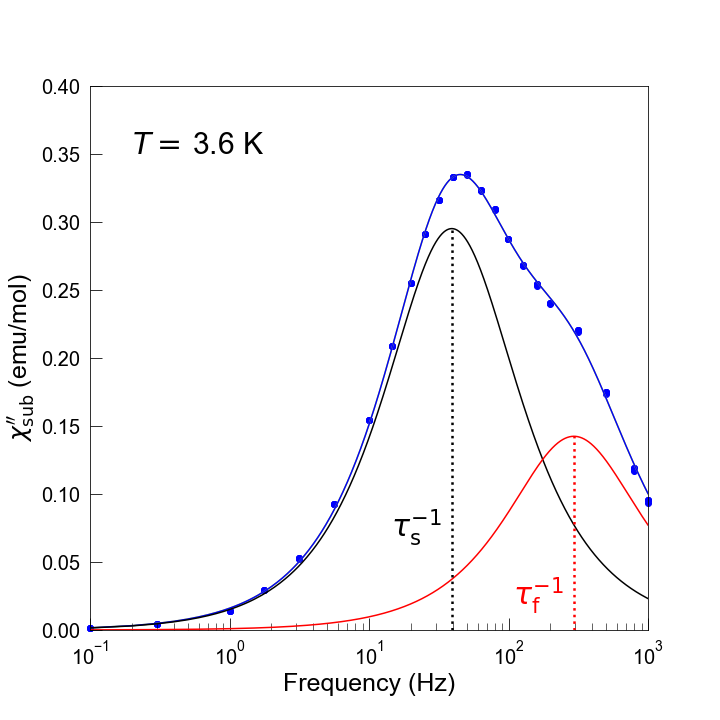}
\caption{(Color online) Frequency dependence of $\chi_{\mathrm{sub}} ^{\prime \prime}$ at 
$3.6\ \mathrm{K}$ and the fitting results by the double Debye relaxation (the blue curve). The black and red curves indicate the slower and faster terms, respectively.}
\label{two_relax}
\end{figure}

   The I-II phase transition at
   $T_\mathrm{N2}$ in
   $\mathrm{DyRu_2Si_2}$ is noteworthy in three points. 
   First, it has the characteristic temperature hysteresis.
   Second, it is accompanied by extremely slow dynamics,
    where the peak anomaly at 
   $T_\mathrm{N2}$ has a significant frequency dependence and 
   it attenuates with increasing frequency. 
   Third, the characteristic frequency of the ``critical dynamics'' of the I-II phase transition
  increases with decreasing temperature, 
    which implies a non-thermally activated origin.

 For further discussion on the dynamics of the I-II phase transition, 
 we subtracted the ``background'' dynamics owing to the phase I, which is observed in the high-frequency range.
  As seen in the 
  $H \mathchar`- T$ phase diagram in Fig. \ref{phase_diag},  
  the phase II appears in the low-field region.
   We, therefore, assumed that the frequency dependence of the ac 
   susceptibility at 
   $H=5\ \mathrm{kOe}$, which characterizes the dynamics of the phase I,
  is the background. 
  Figure \ref{compare} shows the comparison of the frequency dependences of 
  $\chi^{\prime \prime}$ at zero field measured on the cooling process and $\chi^{\prime \prime}$ at 5 kOe.
  It indicates that 
  $\chi ^{\prime \prime}$ at
  $H=5\ \mathrm{kOe}$ is appropriate to be assumed as the background, 
  except for the low-frequency region, where a small broad peak around
  $1\ \mathrm{Hz}$ was observed.
  For the subtraction, first, we neglected the small peak. 
  This is because the peak is considered to originate from the 
  dynamics of the process where the fluctuating spins in the
  $\mathrm{D_I}$ plane are getting aligned to the magnetic field direction as lowering temperature, 
  and such dynamics should be absent at zero magnetic field.
  Thus, as depicted by the pink solid curve in the figure, 
  we presumed the fitting result using 
  $\chi ^{\prime \prime}$ of the high frequency region above 
    $20\ \mathrm{Hz}$ at
  $H=5\ \mathrm{kOe}$ to be the background in the full frequency range. 
  The fitting function is 
  $\chi^{\prime \prime}_\mathrm{bg} (\omega) = \chi_\mathrm{bg} \, \omega\, \tau_\mathrm{bg} / \{1+(\omega\, \tau_\mathrm{bg})^\alpha \}$ , where 
  $\omega$ is the frequency and 
  $\chi_\mathrm{bg}$,
  $\tau_\mathrm{bg}$ and
  $\alpha$ 
  are fitting parameters. 
   This function describes the high-frequency 
   $\chi ^{\prime \prime}$ at 
   $H = 5\ \mathrm{kOe}$
   well, even though it is an ad hoc function and lacks clear physical meaning. 
   \par

\begin{figure}[t]
\vspace*{0.3cm}
\centering
\includegraphics[width=8cm,bb=0 0 720 976]{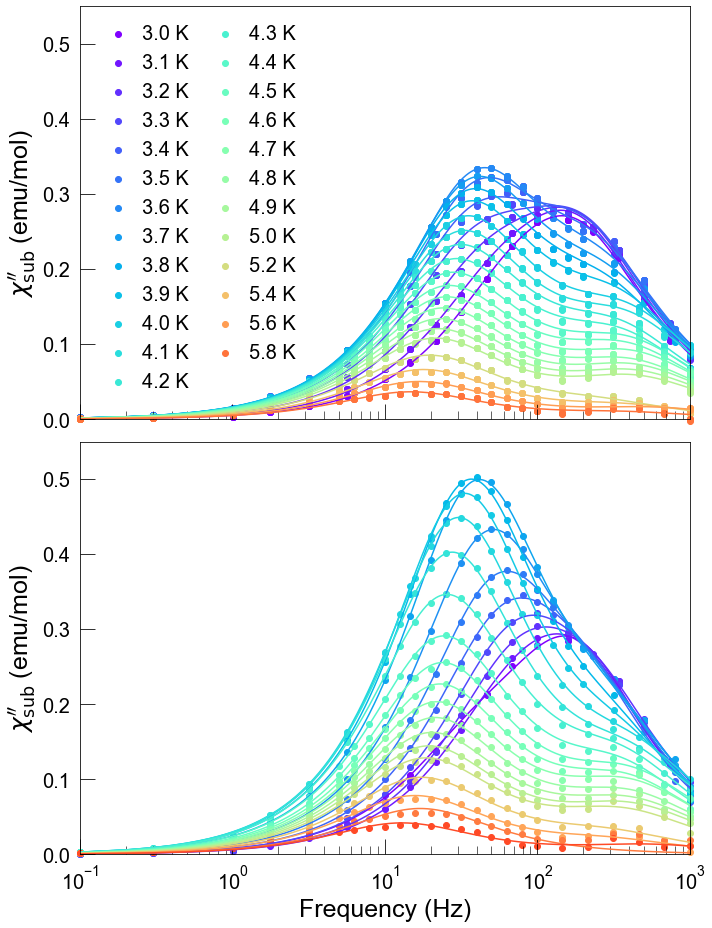}
\caption{(Color online) Frequency dependences of $\chi_{\mathrm{sub}} ^{\prime \prime}$ at various temperatures in the cooling (upper) and heating (lower) processes. The solid lines are the fitting results of the double Debye relaxation.}
\label{chi_sub}
\end{figure}

	\begin{figure*}[!t]
	\begin{tabular}{ccc}
	\centering
		\begin{minipage}{0.33\linewidth}
    		\centering
    		\includegraphics[scale=0.22, bb = 100 0 720 650]{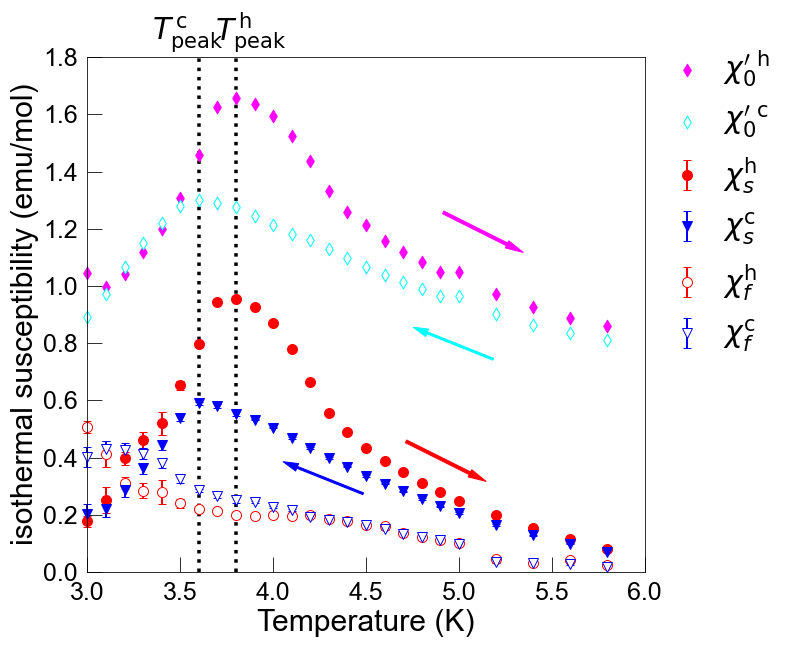}
  		\end{minipage}&
  		
  		\begin{minipage}{0.33\linewidth}
  			\centering
    		\includegraphics[scale=0.22, bb =100 0 720 650]{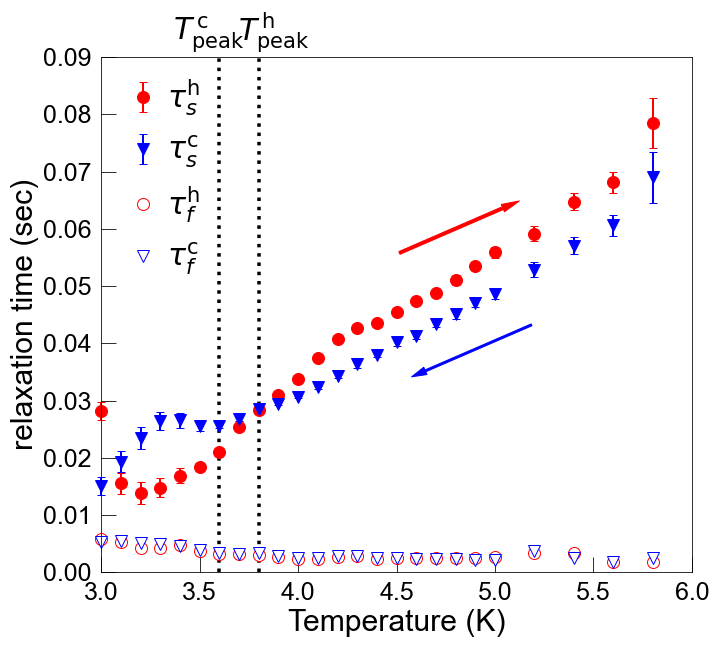}
  		\end{minipage}&
  			
  		\begin{minipage}{0.33\linewidth}
  			\centering
    		\includegraphics[scale=0.22, bb = 100 0 720 650]{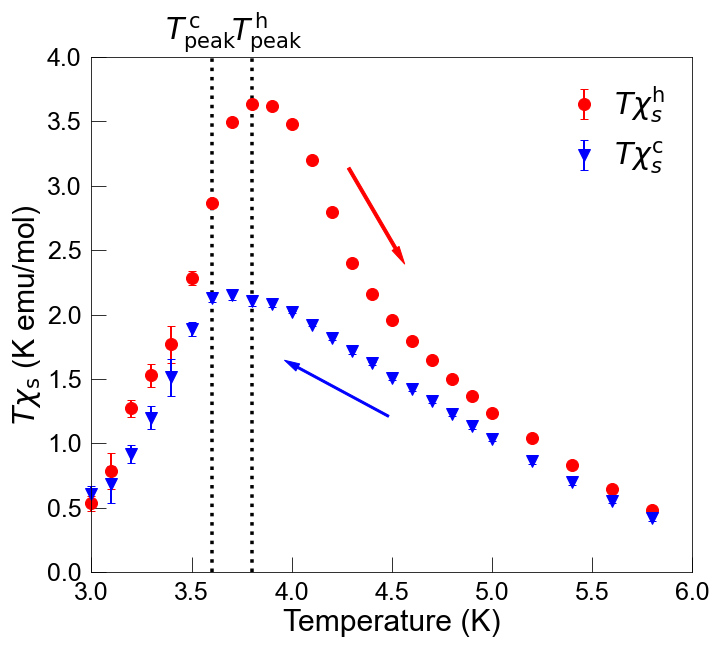}
  		\end{minipage}\\

  		\begin{minipage}{0.33\linewidth}
     		\vspace*{0.3cm}
  			\centering
    		(a)
  		\end{minipage}&
  		
  		\begin{minipage}{0.33\linewidth}
     		\vspace*{0.3cm}
  			\centering
    		(b)
  		\end{minipage}&
  
  		\begin{minipage}{0.33\linewidth}
     		\vspace*{0.3cm}
  			\centering
    		(c)
  		\end{minipage}\\

  	\end{tabular}
  \caption{(Color online) Temperature dependences of the fitting parameters of 
  $\chi_\mathrm{sub} ^{\prime \prime}$ by the double Debye relaxation:
  (a) the isothermal susceptibilities
  $\chi_\mathrm{s}$ and
  $\chi_\mathrm{f}$, and
  (b) relaxation times 
  $\tau_\mathrm{s}$ and 
  $\tau_\mathrm{f}$.
  The real part of the ac susceptibility at 
  $0.1\ \mathrm{Hz}$,
  $\chi_\mathrm{0}^\prime$, is also plotted in (a).
  (c)
  $T\chi_\mathrm{s}$
  as the function of temperature. 
  Superscripts of 
  ``h'' and ``c'' denote the heating and cooling processes, respectively.
   Arrows in the figures represent the directions of the measurement processes.
   }
  \label{fit_results}
\end{figure*}

   Figure \ref{two_relax} shows the frequency dependence of 
   $\chi ^{\prime \prime}$ at 
   $3.6\ \mathrm{K}$ after the background-subtraction.
    Hereafter let this subtracted susceptibility be 
   $\chi_{\mathrm{sub}} ^{\prime \prime}$.
   It has two peak structures around 
   $40\ \mathrm{Hz}$ and
   $300\ \mathrm{Hz}$.
   The frequency dependence of 
    $\chi_{\mathrm{sub}} ^{\prime \prime}$ is well described by double Debye relaxation; 
        
    \begin{equation}
    		\chi ^{\prime \prime}_\mathrm{sub} =  \chi_\mathrm{s} \frac{\omega \tau_\mathrm{s}}{1+ (\omega \tau_\mathrm{s})^2}  
    		+  \chi_\mathrm{f} \frac{\omega \tau_\mathrm{f}}{1+ (\omega \tau_\mathrm{f})^2},	
    \end{equation}
    
  \noindent where 
  $\chi_{\mathrm{s}, \mathrm{f}}$ and 
  $\tau_{\mathrm{s}, \mathrm{f}}$ are the isothermal susceptibilities and relaxation times, and the subscripts ``s'' and ``f'' denote the slower and faster terms with longer and shorter relaxation times, respectively. \par
  
   The temperature variation of the frequency dependence of 
      $\chi_{\mathrm{sub}} ^{\prime \prime}$ in the cooling process is shown in the upper panel of Fig. \ref{chi_sub}.
   Both peaks emerge at 
   $6.0\ \mathrm{K}$, and grow up on cooling. 
   The slower term in the lower frequency region increases towards 
   $T_\mathrm{N2}$ and exhibits maximum with slight shifting towards higher frequency. 
   On the other hand, the faster one keeps on growing on further cooling.
   $\chi ^{\prime \prime}_\mathrm{sub}$ in the heating process was also derived by the same procedure, shown in the lower panel of Fig. \ref{chi_sub}. 
   It is also able to be fit by the double Debye relaxation and shows the similar temperature variation.\par

  Figures \ref{fit_results} (a) and (b) are the temperature dependences of the parameters,
   isothermal susceptibilities and relaxation times, respectively.
  The superscripts of ``h'' and ``c'' denote the parameters in the heating and cooling processes, respectively.
  In Fig. \ref{fit_results} (a), the isothermal susceptibility of the slower term 
  $\chi_\mathrm{s}$ exhibits clear peak and hysteresis around  
  $T_\mathrm{N2}$. 
  $\chi_\mathrm{s}^\mathrm{c}$ and 
  $\chi_\mathrm{s}^\mathrm{h}$ show the peaks at slightly different
  temperatures, 
  $T_\mathrm{peak}^\mathrm{c} = 3.6\ \mathrm{K}$ and 
  $T_\mathrm{peak}^\mathrm{h} = 3.8\ \mathrm{K}$, and the peak in the heating process is more pronounced.
  $\chi_\mathrm{s}^\mathrm{c}$ and 
  $\chi_\mathrm{s}^\mathrm{h}$ merge far above 
  $T_\mathrm{N2}\ (T  >5.6\ \mathrm{K})$ and below 
  $T_\mathrm{N2}\ (T < 3.2\ \mathrm{K})$.
  In the figure, we also plot the real part of the ac susceptibility at the lowest frequency,
  $0.1\ \mathrm{Hz}$,
  $\chi_{0} ^{\prime\,\mathrm{c}}$ and 
  $\chi_{0} ^{\prime\,\mathrm{h}}$ 
  measured on both the cooling and heating processes. 
	They behave similarly to
  $\chi_\mathrm{s}^\mathrm{c,h}$ around
  $T_\mathrm{N2}$. 
  Note that we didn't subtract the background from these susceptibilities 
  $\chi_0^{\prime\, \mathrm{c,h}}$. 
  Thus, it can be concluded that the characteristic features in 
  $\chi_\mathrm{s}^\mathrm{c}$ and 
  $\chi_\mathrm{s}^\mathrm{h}$ are not artifacts owing to the background-subtraction or the phenomenological fitting by the double Debye relaxation. 
  In contrast to the slower term, the isothermal susceptibilities of the faster terms, 
  $\chi_\mathrm{f}^\mathrm{c}$ and
  $\chi_\mathrm{f}^\mathrm{h}$, do not exhibit remarkable hysteresis behaviors and any anomalies around 
  $T_\mathrm{N2}$.
  The both 
  $\chi_\mathrm{f}$ moderately increase down to the lower temperature than 
  $T_\mathrm{N2}$.
  It should be noted that this is consistent with the increase of the dc susceptibility on cooling down to 
  $0.46\ \mathrm{K}$ as shown in the inset of Fig. \ref{T_hys}. 
  The difference between the temperature dependences of 
  $\chi_\mathrm{s}$ and 
  $\chi_\mathrm{f}$ obviously indicates that the slower and faster terms of 
  $\chi_\mathrm{sub} ^{\prime \prime}$ are attributed to different dynamics: 
  the critical dynamics of the I-II phase transition and the dynamics of the disordered spins in the phase II, respectively.

  Figure \ref{fit_results} (b) shows the temperature dependences of the
   relaxation times of the slower and faster components, 
   $\tau_\mathrm{s}$ and
   $\tau_\mathrm{f}$. 
   The relaxation time  
   $\tau_\mathrm{s}$, which is attributed to the dynamics of I-II phase transition, reveals two noteworthy facts about it.
   First, the relaxation time is extremely long.
   At around
   $5.8\ \mathrm{K}$, it is in the order of 
   $100\ \mathrm{ms}$ and
  it declines almost linearly towards
   $T_\mathrm{N2}$.
  Second, it is indicated that the dynamics of the I-II phase transition is non-thermally activated, because the relaxation time
   decreases with temperature decreasing. 
  This is consistent with the temperature variation of 
  $\chi ^{\prime \prime} _\mathrm{sub}$ shown in Fig. \ref{chi_sub}
   and it is more apparent here.
	It should be also noted that the critical slowing down towards
	$T_\mathrm{N2}$ is absent. As seen in the isothermal susceptibility 
	$\chi_\mathrm{s}$, the
	relaxation time 
	$\tau_\mathrm{s}$ also shows the hysteresis behavior,
	especially below 
	$T_\mathrm{N2}$.   
	The relaxation time in the cooling process 
	$\tau_\mathrm{s}^\mathrm{c}$ shows a hump at around 
	$3.4\ \mathrm{K}$.
	On the other hand, that in the heating process
   $\tau_\mathrm{s}^{\mathrm{h}}$ shows a minimum at around 
   $3.2\ \mathrm{K}$ and is smaller than
   $\tau_\mathrm{s}^{\mathrm{c}}$. 
   Above 
   $T_\mathrm{N2}$, the size relationship between 
   $\tau_\mathrm{s}^\mathrm{c}$ and 
   $\tau_\mathrm{s}^\mathrm{h}$ is reversed, namely, 
	$\tau_\mathrm{s}^\mathrm{h}$ is slightly larger than 
   $\tau_\mathrm{s}^\mathrm{c}$. 
   In contrast to 
   $\tau_\mathrm{s}$, the relaxation time of the faster term
   $\tau_\mathrm{f}$ doesn't show significant hysteresis behavior and 
   increases moderately with temperature decreasing below 
    $T_\mathrm{peak}^\mathrm{c}$. 
    It indicates that the faster dynamics is thermally activated.
  Again, these differences between 
  $\tau_\mathrm{s}$ and
  $\tau_\mathrm{f}$ indicate the different origins of the two dynamics.

   Figure \ref{fit_results} (c) shows the temperature dependences of products of the isothermal susceptibility and temperature 
  $T \chi_\mathrm{s}$ for the two processes, which corresponds to the spin correlations;
  
	\begin{equation}
	T \chi_\mathrm{s} \propto \sum_{i,j}<S_i \, S_j>.
    \end{equation}

  \noindent This quantity increases when ferromagnetic (FM) spin correlations develop, 
  whereas, decreases when antiferromagnetic (AFM) spin correlations develop.
  The I-II phase transition is the process where the spins in the disordered 
   $\mathrm{D_I}$ plane form the striped AFM order along the b-axis. 
   Nevertheless, 
   $T \chi_\mathrm{s}^\mathrm{c,h}$ increase towards 
   $T_\mathrm{N2}$, which indicates that 
   growth of dynamic FM correlations is involved in the phase transition.
   It might look contradictory, however, as discussed in Sec. 5, it indicates 
   important information about this extraordinary phase transition. 

     \section{Discussion}

	In this section, we propose the process of the I-II phase transition indicated by the above analysis.

	 As a summary of the analysis, the features of the dynamics attributed to the I-II phase transition are the following;
	\begin{enumerate}
	     \item Dynamic FM correlations with long relaxation time appear at around
				 $6\ \mathrm{K}$ and grow towards the I-II phase transition temperature 
				  $T_\mathrm{N2}$,
		 \item The dynamics is non-thermally activated, 
		 \item  The dynamics shows hysteresis behavior as shown in the temperature dependences of 
				 $\chi_\mathrm{s}$ and
				 $\tau_\mathrm{s}$ (Figs. \ref{fit_results} (a) and (b)). 
	\end{enumerate}
	
	  The feature (1) indicates that large dynamic FM spin textures appear precedently to the phase transition.
	 In general, the response time of  a spin system is in the order of ps-ns. 
	 Thus, the observed long relaxation time indicates that the spin textures should be considerably large.
	 Since the I-II phase transition is the process where the fluctuating spins in the
	 $\mathrm{D_I}$ planes form the striped order of the phase II, it is reasonable to consider that the FM spin textures appear and grow in the 
	 $\mathrm{D_I}$ plane and they are strongly involved with the development of the striped spin correlations. \par

\begin{figure*}[!t]
			\vspace*{-7.5cm}
			\begin{center}
				\begin{minipage}{\linewidth}
				\centering
				\includegraphics[width=1.9\linewidth,bb = 0 0 1979 950]{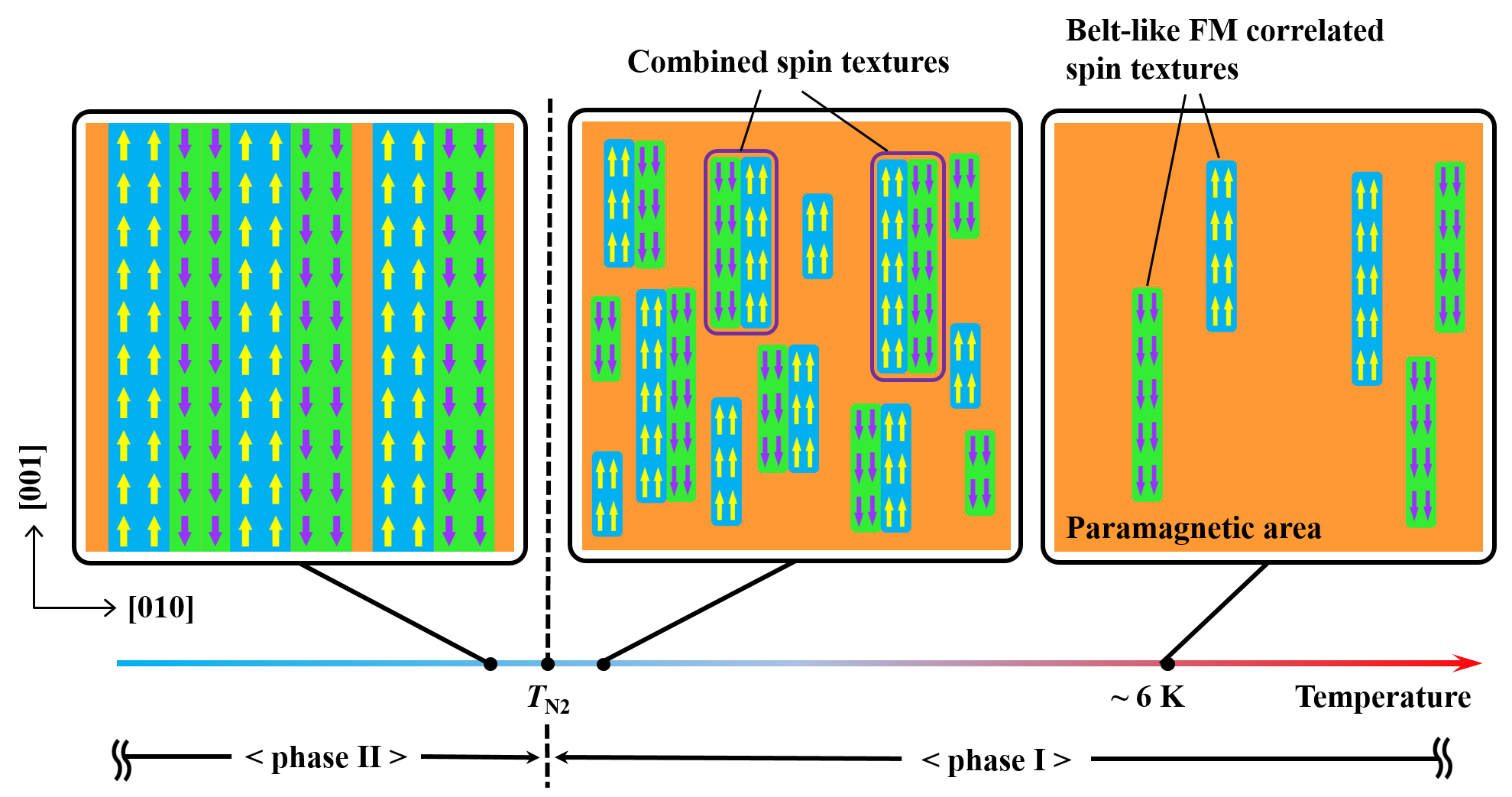}
				\end{minipage}
			\end{center}
			\caption{(Color online) Schematic picture of the ordering process of the I-II phase transition in the
			$\mathrm{D_I}$ plane, which is represented by the column of the orange cross-squares surrounded by the green dashed rectangle in Fig. \ref{magn_struc}. 
			All the panels show the projection of the $\mathrm{D_I}$ plane onto the a-plane.
			The orange-shaded areas represent the region where the spins are fluctuating (noted 
			as paramagnetic area) and the blue and green rectangles represent the ``belt-like'' FM spin textures 
			whose spins are parallel and antiparallel with the c-axis. 
			The purple rectangles in the middle panel represent the antiparallelly combined belt-like spin textures.}
			\label{phase_trans_mecha}
		\end{figure*}

	  On the basis of these considerations, we propose a schematic picture of the I-II phase transition as shown in 
	  Fig. \ref{phase_trans_mecha}, 
	  where the development and shift of spin correlations with temperature variation in the
	  $\mathrm{D_I}$ plane, which is represented by the column of the orange cross-squares surrounded by the green dashed rectangle in Fig. \ref{magn_struc}, are shown.
		All the panels show the projection onto the a-plane. 
	  The striped pattern of the phase II in Fig. \ref{phase_trans_mecha} indicates that the nearest-neighbor (NN) interaction is FM along both the b- and c-axes and the AFM next-nearest-neighbor (NNN), and maybe further long-range interactions, compete with the NN FM interaction along the b-axis, whereas, further interactions are too weak to compete with the NN FM interactions along the c-axis.  
	  Thus, we hypothesize that the non-frustrated two-row FM spin textures are formable at much higher temperature 
	  than the phase transition temperature and the large ``belt-like'' FM correlated spin textures emerge around
	  $6\ \mathrm{K}$
	  as precursors of the striped AFM structure of the phase II, as shown in the right panel of Fig. \ref{phase_trans_mecha}.
	  Then, they grow larger and become denser with decreasing temperature, 
	  and finally spontaneously form the striped magnetic structure at
	  $T_\mathrm{N2}$ as shown in the left panel of Fig. \ref{phase_trans_mecha}.

	  The non-thermally activated behavior of the relaxation time
	  $\tau_\mathrm{s}$ can be explained in this picture as follows. 
	  The key of the explanation is that the relaxation time reflects the size of the spin textures,
	  namely, larger spin textures fluctuate more slowly and have a longer life span and smaller ones have shorter lifetimes.
	   As we discuss above, when decreasing temperature, the precedently emergent belt-like spin textures grow up and become denser, and then, they are
	   combined antiparallel to each other along the b-axis by the AFM NNN and further long-range interactions.
	   It is schematically shown in the middle panel of Fig. \ref{phase_trans_mecha} by the purple rectangle frame.
	   Once they are combined, the spin textures are not FM but striped AFM, thus they no longer contribute to the response to the uniform ac field
	   dominantly. 
	   On the other hand, the remaining small fragment of FM spin textures predominantly contribute to the ac response.
	   As a result, the relaxation time of the dynamics measured by the ac susceptibility becomes shorter as approaching $T_\mathrm{N2}$ because the size of the FM spin textures predominantly responding to the ac field becomes smaller. That is to say, the non-thermally activated behavior of the ac response may indicate the shift of the spin correlations from the isolated FM spin textures to the striped AFM ones. 
\par

   The fact that there is the temperature hysteresis over the I-II phase transition can be also explained. 
   In this picture, the I-II phase transition is the spontaneous arrangement of the large belt-like FM spin textures. 
   This ordering process should be extremely slower than conventional magnetic phase transitions because the time-scale of the dynamics of the units of the ordering, 
   the belt-like FM spin textures, is very long, being the order of 10-100 ms. 
   Thus, the temperature hysteresis should be observed in the conventional experimental time-scale (order of second or minute). 
   The details of the hysteresis behaviors observed in 
   $T \chi_\mathrm{s}$ and
   $\tau_\mathrm{s}$ are discussed below.     
   As shown in Figs. \ref{fit_results} (c) and (b),
    $T \chi_\mathrm{s}$  in the heating process is larger than that in the cooling one, and
    $\tau_\mathrm{s}$ in the heating process is shorter than that in the cooling one below
    $T_\mathrm{N2}$ and is longer above 
    $T_\mathrm{N2}$.
    In our scenario, these hysteresis behaviors can be interpreted as the difference between the ``solidification'' of the belt-like FM spin textures 
    and ``melting'' of the striped order in the
    $\mathrm{D_I}$ plane, 
    which correspond to the cooling and heating processes, respectively.
    The former is the process described so far 
    and the latter is the opposite process where the complete striped structure of the phase II 
    breaks into the belt-like FM textures. 
    Above 
    $T_\mathrm{N2}$, it is naturally expected that larger FM spin textures are more densely
    persist in the ``melting'' process and, vice versa, smaller spin textures should be frequently formed
    in the ``solidification'' process.
    Thus, the spin correlations
    $T \chi_\mathrm{s}$ should be larger on the heating process than those on the cooling one and the relaxation time
    $\tau_\mathrm{s}$ in the heating process should be longer than that in the cooling process above 
    $T_\mathrm{N2}$.
    On the other hand, since the stable state is the stripe AFM below 
    $T_\mathrm{N2}$, the size relationship of the fragment of FM spin textures is reversed,
     namely, larger spin textures persist in the solidification process and vice versa. 
    Thus, 
    $\tau_\mathrm{s}$ below 
    $T_\mathrm{N2}$ should be longer in the cooling process than that in the heating process.
        
    At last, we discuss the question that the novel slow critical dynamics is only present in the I-II phase transition but is
    absent in the para-I phase transition at 
    $T_\mathrm{N1}$. 
    Probably, this comes from the difference between the dimensionality of these two phase transitions.
    As we mentioned in Sec. 1, paramagnetic spins in the
    $\mathrm{D_I}$ plane can be considered as a pseudo two-dimensional system.
    The I-II phase transition is the process where the fluctuating spins in these
    $\mathrm{D}_\mathrm{I}$ planes order into the striped structure.
    On the other hand, the para-I phase transition is the process where the fluctuating spins in the three-dimensional paramagnetic state order.
    In general, low dimensionality enhances the fluctuations and destabilizes the ordered state in the system. 
    The extraordinary ordering process of the I-II phase transition, namely, the precedent emergence of the large spin textures 
    and their spontaneous arrangement can be a consequence of its low dimensionality.
   
\section{Summary}
	We performed the ac susceptibility measurements of the frustrated magnet 
	$\mathrm{DyRu_2Si_2}$, especially in the vicinity of the phase transition between the partially ordered phases I and II.
	Detailed analysis of the temperature and frequency dependences of the ac susceptibility reveals the novel critical
	dynamics of the I-II phase transition.
    The temperature dependences of the relaxation time and isothermal susceptibility indicate the following three striking features.
	First, the dynamic FM correlations with extremely long relaxation time appear precedently at around 
	$6.0\ \mathrm{K}$ and grow towards the phase transition temperature
	$T_\mathrm{N2}$.
	Second, the dynamic FM correlations exhibit non-thermally activated behavior.
	Third, the dynamics shows the hysteresis behavior.
	On the basis of these features, we propose the process of this phase transition as shortly described as follows.
   In the 
  $\mathrm{D_I}$ plane, which is the emergent two-dimensional disordered system in the phase I, 
  the large and stable belt-like FM spin textures formed by the NN FM interactions appear as precursors around 
  $6.0\ \mathrm{K}$.
  With decreasing temperature towards
  $T_\mathrm{N2}$, they become denser, more likely to come next to each other along the b-axis and combined when they are antiparallel 
  to each other due to the NNN AFM and further long-range interactions.
  Eventually, they spontaneously form the striped structure of the phase II at 
  $T_\mathrm{N2}$. 
  \par
   
   We will perform the neutron scattering experiments in the near future to verify our hypothesis about the ordering process of
  the I-II phase transition.
  We expect to observe the development and shift of the spin correlations from the broad FM correlations to the sharp stripe ones around
   $(2/9, 2/9, 0)$ in the reciprocal lattice space, which substantially verifies our hypothesis on the spin ordering in the pseudo two-dimensional plane. 
  
\ 

{\bf Acknowledgements}

The authors acknowledge the support from the JSPS Grant-in-Aid for Scientific Research (B) (No. 20H01852).

\bibliographystyle{jpsj}

\end{document}